\documentclass[reprint, prb, superscriptaddress, floatfix]{revtex4-1}

\usepackage{savesym}
\usepackage{hyperref}
\usepackage{graphicx}
\usepackage{amsmath}

\savesymbol{comment}
\usepackage{wasysym}
\usepackage{amsfonts}
\usepackage{color}
\usepackage{verbatim}
\usepackage{upgreek}
\usepackage{bm}
\usepackage{units}
\usepackage{placeins}
\usepackage{babel}
\restoresymbol{TXF}{comment}
\usepackage{float}
\usepackage{tabularx}

\usepackage[draft]{changes}
\newcommand{\beq}{\begin{equation}}
\newcommand{\eeq}{\end{equation}}
\newcommand{\bea}{\begin{eqnarray}}
\newcommand{\eea}{\end{eqnarray}}
\newcommand{\bal}{\begin{align}}
\newcommand{\eal}{\end{align}}

\newcolumntype{L}[1]{>{\raggedright\arraybackslash}p{#1}}
 
\newcolumntype{C}[1]{>{\centering\arraybackslash}p{#1}}
 
\newcolumntype{R}[1]{>{\raggedleft\arraybackslash}p{#1}}

\begin{document}
\title{Ultrafast lattice dynamics and electron-phonon coupling in platinum extracted with a global fitting approach for time-resolved polycrystalline diffraction data}
\author{Daniela Zahn} 
 \email{zahn@fhi-berlin.mpg.de}
\affiliation{Fritz-Haber-Institut der Max-Planck-Gesellschaft, Faradayweg 4-6, 14195 Berlin, Germany}

\author{H\'{e}l\`{e}ne Seiler}
\affiliation{Fritz-Haber-Institut der Max-Planck-Gesellschaft, Faradayweg 4-6, 14195 Berlin, Germany}

\author{Yoav William Windsor}
\affiliation{Fritz-Haber-Institut der Max-Planck-Gesellschaft, Faradayweg 4-6, 14195 Berlin, Germany}
\affiliation{Institut für Optik und Atomare Physik, Technische Universit\"{a}t Berlin, Stra\ss e des 17. Juni 135, 10623 Berlin, Germany}

\author{Ralph Ernstorfer}
 \email{ernstorfer@fhi-berlin.mpg.de}
\affiliation{Fritz-Haber-Institut der Max-Planck-Gesellschaft, Faradayweg 4-6, 14195 Berlin, Germany}
\affiliation{Institut für Optik und Atomare Physik, Technische Universit\"{a}t Berlin, Stra\ss e des 17. Juni 135, 10623 Berlin, Germany}

\begin{abstract}

Quantitative knowledge of electron-phonon coupling is important for many applications as well as for the fundamental understanding of nonequilibrium relaxation processes. Time-resolved diffraction provides direct access to this knowledge through its sensitivity to laser-induced lattice dynamics. Here, we present an approach for analyzing time-resolved polycrystalline diffraction data. A two-step routine is used to minimize the number of time-dependent fit parameters. The lattice dynamics are extracted by finding the best fit to the full transient diffraction pattern rather than by analyzing transient changes of individual Debye-Scherrer rings. We apply this approach to platinum, an important component of novel photocatalytic and spintronic applications, for which a large variation of literature values exists for the electron-phonon coupling parameter $G_\mathrm{ep}$. Based on the extracted evolution of the atomic mean squared displacement (MSD) and using a two-temperature model (TTM), we obtain $G_\mathrm{ep}=\unit[(3.9\pm0.2)\times10^{17}]{\frac{W}{m^3K}}$ (statistical error). We find that at least up to an absorbed energy density of \unit[124]{J/cm$^3$}, $G_\mathrm{ep}$ is not fluence-dependent. Our results for the lattice dynamics of platinum provide insights into electron-phonon coupling and phonon thermalization and constitute a basis for quantitative descriptions of platinum-based heterostructures in nonequilibrium conditions.

\end{abstract}

\maketitle
\date{\today}

\section{Introduction}
Platinum is important to many technological fields, and in particular it is often used as a catalyst. Many chemical reactions require high temperatures and/or pressures, rendering them energy-intensive. A promising approach to reduce this energy cost is to employ photocatalysis. One emerging approach is to employ bimetallic heterostructures that combine a catalytically active material (e.g. platinum or palladium) with a plasmonic material (e.g. silver, gold or aluminum)~\cite{Sytwu2019}, for example, in so-called antenna-reactor nanostructures~\cite{Robatjazi2020,Zhang2016}. Another approach is to combine a catalytically active metal with a semiconductor (semiconductor-metal heterostructure)~\cite{2018Bai}. It has been demonstrated that such structures can exhibit high photocatalytic activities~\cite{Zhang2016,Robatjazi2020,2018Bai}. In many photocatalytic reactions, highly excited ("hot") electrons play the decisive role~\cite{2018Zhang,Zhou2018,2002Lei}. The time scales on which the electrons remain hot depends on their coupling to other degrees of freedom, in particular the lattice, which is the main heat sink on ultrafast time scales. Therefore, knowledge about the ultrafast lattice dynamics is important for understanding the dynamics of hot-carrier driven chemical reactions.

In addition to its use in catalysis, platinum is also an important material for spintronics due to its large spin-orbit coupling. This leads, for example, to a large Spin-Hall effect, which is widely employed for spin-to-charge or charge-to-spin conversion~\cite{Uchida2008,2015Kehlberger,Demidov2012,Seifert2016,Yan2017,Jungwirth2012}, and to a large Rashba effect~\cite{Miron2011,Miron2010}. The functionality of spintronic heterostructures is determined by the interplay of interfacial couplings and couplings within the individual materials. The relaxation processes within a material can therefore strongly influence charge and spin currents across interfaces. For example, the spin-Seebeck current across a photoexcited yttrium iron garnet/platinum interface strongly depends on the electronic temperature in platinum~\cite{Seifert2018}, whose evolution is dominated by electron-lattice equilibration. Therefore, knowledge of the lattice response of platinum serves as a basis for understanding and controlling the behavior of spintronic devices.

In particular, it is of interest to quantify the coupling of excited carriers to the lattice. However, literature values for the electron-phonon coupling constant $G_\mathrm{ep}$ of platinum vary significantly~\cite{Jang2020,2000Hohlfeld,Caffrey2005,Choi2015,2002Lei,2020Medvedev,2020Smirnov}, from $\unit[{\raise.17ex\hbox{$\scriptstyle\sim$}}\hspace{1pt}0.45\times10^{17}]{\frac{W}{m^3K}}$ (room-temperature value)~\cite{2020Medvedev} to $\unit[(10.9\pm 0.5)\times10^{17}]{\frac{W}{m^3K}}$~\cite{Caffrey2005}. So far, most experimental values for $G_\mathrm{ep}$ in platinum were deduced using optical methods, for example time-resolved optical reflectivity (TRR) measurements~\cite{Caffrey2005,2000Hohlfeld,Choi2015,Kimling2017}. A challenge of TRR measurements is that the reflectivity change depends on both the electron and the phonon temperatures~\cite{1972Rosei,1990Brorson,2008Carpene}, and separating these contributions is non-trivial. In addition, the dependence of the reflectivity change on these temperatures is not always linear, especially for higher fluences~\cite{Norris2003,Smith2001} and for transition metals~\cite{2000Hohlfeld}. An alternative way to access $G_\mathrm{ep}$ is by using a ferromagnetic detection layer in combination with time-resolved MOKE measurements~\cite{Jang2020}. However, this approach relies on modeling the nonequilibrium responses of both platinum as well as of the detection layer, which limits the precision of the method. In contrast, time-resolved photoemission spectroscopy provides direct access to transient electron temperatures. Determination of $G_\mathrm{ep}$ with this surface-sensitive technique remains challenging, however, due to the interplay of electron relaxation and transport. Therefore, time-resolved diffraction is an important complementary technique to optical and photoemission measurements. It is only sensitive to the lattice and can therefore directly measure the lattice response to photoexcitation, thus providing quantitative insights into electron-lattice equilibration from the lattice perspective. However, quantitatively extracting the lattice dynamics from time-resolved diffraction data is not trivial. This is because laser excitation causes the intensity of the Bragg reflections to decrease due to the Debye-Waller effect, but it simultaneously enhances thermal diffuse scattering, which contributes to the background. Separating these effects is particularly challenging for polycrystalline samples compared to single crystals, because the diffraction signal is weaker and the two effects overlap in the probed momenta. An additional challenge is the limited transverse coherence in time-resolved electron diffraction experiments, which can cause significant overlap of diffraction rings. As a consequence, when the transient changes of each diffraction ring are analyzed separately using a fit, fit parameters of adjacent rings can strongly correlate. In addition, extracting the intensity changes of each ring separately can lead to inconsistencies, since different rings can yield different results for the amplitudes of the atomic mean-squared displacement (MSD) change. Therefore, in such cases, an approach which extracts the lattice dynamics based on the full diffraction pattern and minimizes the number of time-dependent fit parameters is desired.

Here, we present such an approach for the analysis of polycrystalline diffraction patterns, which consists of two steps. We use the term "global" to describe our approach since the full diffraction pattern is taken into account in the analysis and there is only one fit parameter for the MSD change and one fit parameter for the lattice expansion. The latter reduces the number of time-dependent fit parameters compared to analyses in which the intensity and position of each diffraction ring are time-dependent fit parameters. We apply this global approach to femtosecond electron diffraction data of platinum.
Next, we convert the results for the MSD  change into lattice temperature, perform a fit to a two-temperature model~\cite{1974Anisimov,1987Allen} (TTM) and extract a value for the electron-phonon coupling parameter. We also discuss the role of non-thermal phonons in the lattice dynamics. Our results provide quantitative information about electron-lattice equilibration as well as phonon thermalization, which are integral to the understanding of nonequilibrium dynamics in platinum.

\section{Experiment}
\begin{figure}[p]
\begin{center}
\includegraphics[width=\columnwidth]{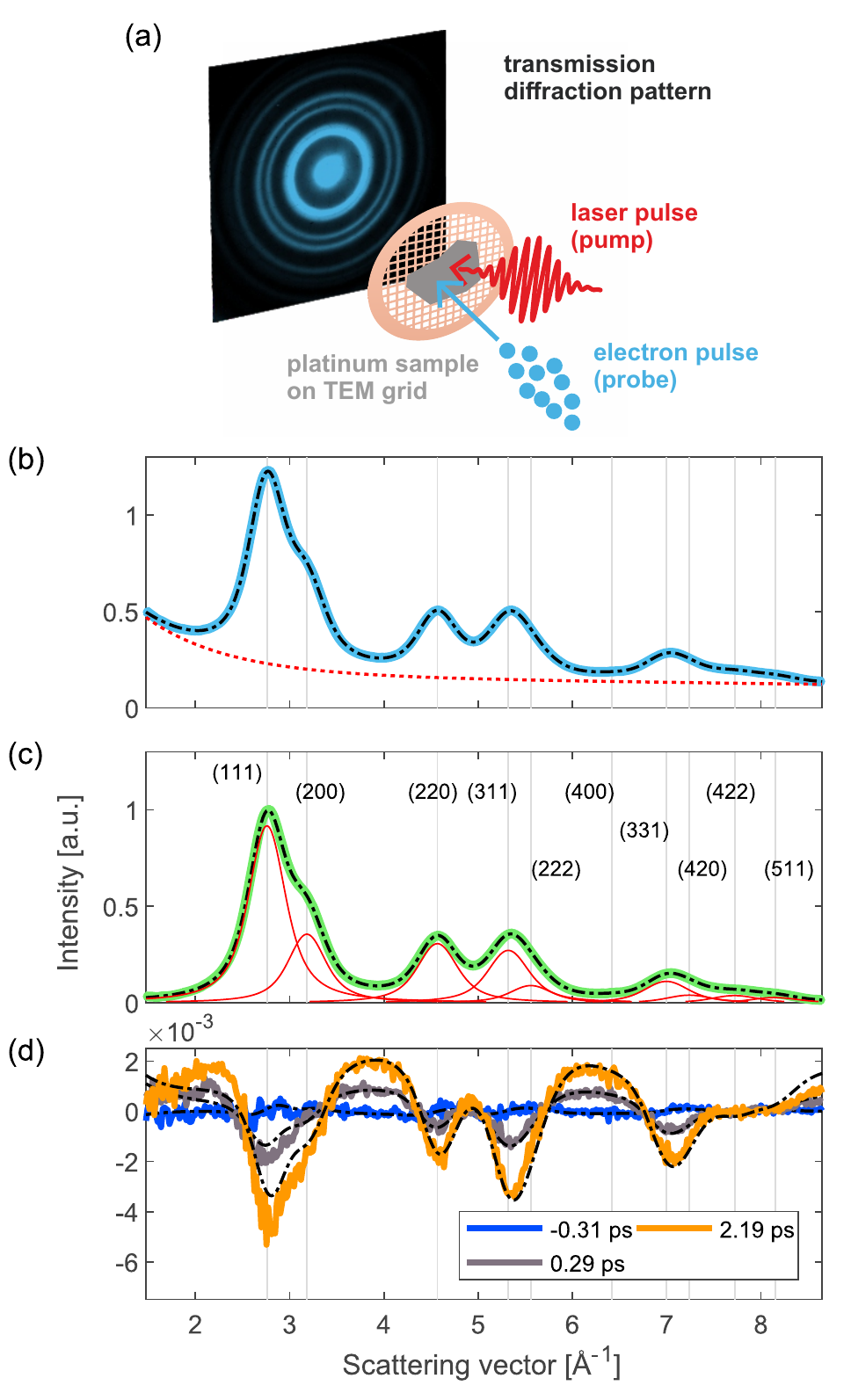}
\caption{Experimental setup and diffraction pattern of platinum. (a) Schematic illustration of the electron diffraction experiment. The freestanding thin-film sample is excited with an ultrashort laser pulse and the lattice response is probed with an ultrashort electron pulse. Diffraction patterns for several pump-probe delays are recorded in transmission. (b) Azimuthally averaged diffraction pattern (radial profile, RP) of our platinum sample. The solid blue curve corresponds to the experimental data and the dashed black curve represents the static global fit  (see text for details about the fit). The dashed red curve shows the static fit result for the background contribution. (c) RP after subtracting the fit result for the background. The solid green curve shows the experimental data and the dashed black curve shows the static fit result without the background contribution. The contributions of the individual Bragg reflections are displayed in red. (d) Differences between the RP's for selected delays after photoexcitation and the RP prior to photoexcitation. The experimental data are shown as solid curves and the fit results of the dynamic fit are shown as dashed black curves. Note that in Panels~(b)-(d), the x-axis was converted from image pixels to scattering vectors using the results of the static fit (for illustration purposes).}
\label{fig:1}
\end{center}
\end{figure}

We study the lattice dynamics of platinum using the compact femtosecond electron diffractometer described in Ref.~\cite{2015Waldecker}. The sample is a freestanding, polycrystalline film of platinum with a thickness of \unit[15]{nm}. It was deposited on a NaCl single crystal using electron beam evaporation. Then, the platinum film was transferred to a standard TEM grid using the floating technique~\cite{2007Dwyer}. To excite the sample, we use infrared laser pulses with a photon energy of \unit[0.70]{eV} and a pulse length of around \unit[80]{fs} (FWHM). The lattice response to photoexcitation is probed using ultrashort electron pulses with a kinetic energy of \unit[70]{keV}. The estimated time resolution of the experiment is around \unit[170]{fs}. The electrons are diffracted by the sample and diffraction patterns are recorded in transmission. Figure~\ref{fig:1}(a) shows a schematic illustration of the electron diffraction experiment, including a diffraction image of our platinum sample. 

Since our sample is polycrystalline, we observe Debye-Scherrer rings. Therefore, we azimuthally average the diffraction images for further analysis. In addition, the images are background-subtracted and flatfield-corrected and the (delay-averaged) azimuthal averages are normalized by dividing by the average image intensity (excluding the zero order beam and regions outside the field of view). We choose this normalization method as opposed to normalization to the lowest order diffraction peak to avoid systematic errors due to multiple scattering.

The resulting radial profile (RP) is displayed in Fig.~\ref{fig:1}(b). Each radial profile corresponds to the average of 20 scans of the pump-probe delay, except for the lowest excitation density, for which 24 scans were recorded. For every point of a pump-probe scan, 13 diffraction images with an integration time of \unit[5]{s} each were recorded and averaged, corresponding to $13\times5\times4000=260000$ electron pulses in total.

Our main observable is the intensity of the diffraction rings, which weaken as the atomic MSD increases when additional phonons are created (Debye-Waller effect). The relationship between the intensity decrease and the MSD is given by~\cite{Peng}:
    \begin{equation}
        \frac{I(t)}{I_0}=\mathrm{exp}\{ -\frac{1}{3}\hspace{2pt}q^2\hspace{2pt}\Delta \langle  u^2\rangle (t)\}
        \label{eq:Debye-Waller}
    \end{equation}
Here, $\Delta \langle u^2\rangle(t)$ is the MSD change, $I_0$ is the intensity before laser excitation and $q$ is the scattering vector of the diffraction ring, $q=4\pi\hspace{1pt}\mathrm{sin}(\theta)/\lambda$. This description can only be applied to intensity changes due to incoherent phonons, for instance generated by electron-phonon scattering. It can be used to describe the effects of both non-thermal and thermal phonon distributions~\cite{SCOPS,2017Waldecker,BP1}.

\section{Global diffraction data analysis}
To analyze the time-resolved polycrystalline diffraction data, we utilize a fit routine in which the MSD is extracted using the whole RP, instead of specific Debye-Scherrer rings. A major advantage of this approach is that the total number of fit parameters can be reduced significantly, especially for the description of the laser-induced changes in the diffraction pattern. This is possible since some information is encoded in all diffraction rings, namely lattice expansion and MSD increase. Therefore, a global fit can avoid artifacts for example due to strong correlations between fit parameters describing overlapping diffraction rings. Disentangling the dynamics of overlapping rings is particularly important in time-resolved electron diffraction due to the finite coherence of the pulsed electron sources. In particular, there is a trade-off between coherence, time resolution and signal-to-noise ratio, because a smaller source size increases space-charge effects. In the following, we provide more details about our global fit routine, which consists of two steps: In the first step (static fit), we perform a fit to the RP's before laser excitation. In the second step (dynamic fit), a fit is performed to describe the changes of the RP's following laser excitation. Here, we only include changes due to an increase of the phonon population, hence we do not include any photo-induced rearrangement of atoms within the unit cell, e.g. structural phase transitions. The second step of our fitting routine yields the MSD change, the lattice expansion (i.e., the changes in diffraction ring radii) and the background change as a function of pump-probe delay.

\subsection{Static fit}

In the first step of the fitting routine, the goal is to accurately describe the RP's before laser excitation. For this step, first we average all RP's recorded before the arrival of the pump pulse in order to obtain maximum signal-to-noise ratio. For the fit, we use a wide range of reciprocal space that includes the first ten diffraction rings of platinum, from the (111) to the (511) reflection, as shown in Fig.~\ref{fig:1}(c). Despite the low intensity of the diffraction rings at scattering vectors above $\unit[6]{Å^{-1}}$, including these rings in the analyzed range is beneficial as they feature intensity reductions similar to those of lower order peaks due to a large Debye-Waller effect, as shown in Fig.~\ref{fig:1}(d).

Our fit function consists of the sum of a background function and peak functions for the azimuthally averaged diffraction rings. The possibility of fitting all diffraction rings and a global background together in the same step is an additional advantage of the global fit, because it yields a reliable background determination. This becomes important in the second step of the fitting routine (dynamic fit), because the background subtraction strongly influences the results for the amplitude of the MSD increase.

The positions of the diffraction peaks in reciprocal space are known. Thus, in order to reduce free parameters in the fit, we don't consider the diffraction ring radii as independent variables. However, in practice, our magnetic lens used to focus the electron beam on the detector can introduce image distortions, which cause a non-linear relationship between scattering vector and pixel number in the RP's (radial distortion). To account for these distortions, we introduce a correction parameter $\gamma$ as a fit parameter. The radius of the first ring, $r_1$, is also a fit parameter and accounts for the conversion of scattering vectors to image pixels. The other ring radii are then given by:
\begin{equation}
r_{i}=\left( \frac{q_i}{q_1} \right)^\gamma\times r_{1}
    \label{eq:positions_approx}
\end{equation}
Here, $q_i$ are the scattering vectors of the diffraction rings. In our case, the magnetic lens introduces a barrel distortion, and therefore the fit result for  
$\gamma$ is a value slightly smaller than one (here: around 0.96). We expect that our distortion correction can also correct pincushion distortions ($\gamma>1$). Note that only radial distortions can be corrected with $\gamma$.

With Eq. \ref{eq:positions_approx}, the approximate ring radii can be well described. In addition, to refine the ring radii, we introduced individual radius correction factors $f_\mathrm{i}$ as fit parameters. These were constrained such that the radii couldn't deviate more than 2\%  from the values given by Eq.~\ref{eq:positions_approx}. We found that for our experimental data, introducing the individual correction factors $f_i$ in addition to the global distortion correction $\gamma$ is essential to obtain good agreement to the diffraction pattern before laser excitation, which then also determines the quality of the dynamic fit. However, most of the distortion is corrected by $\gamma$, which minimizes erroneous peak positions in the static fit result caused by strong correlations of amplitude and position parameters of overlapping diffraction rings. Note that $\gamma$ and $f_i$ are parameters related to the measurement system. In the absence of distortions, or if distortions were corrected previously, the correction factors $\gamma$ and $f_i$ are not necessary and can be set to 1.

The azimuthally averaged diffraction rings are described as Lorentzians with their amplitudes being unconstrained fit parameters. This makes the fitting procedure suitable also for polycrystalline samples with a preferred orientation, which exhibit different relative diffraction ring intensities compared to powder diffraction patterns.
The width of the Lorentzians is one fit parameter ($w_\mathrm{reci}$), such that the widths of all peaks in reciprocal space are the same. Due to the lens distortions, the widths in pixels, $w_\mathrm{i}(w_\mathrm{reci},r)$, are slightly different for the different rings. We don't consider broadening due to finite crystallite sizes according to the Scherrer equation, because the width of the rings is dominated by the finite coherence of the pulsed electron beam. In total, the function describing the diffraction rings is given by:
\begin{equation}
    F(r)=\sum_{i=1}^{N}\frac{A_i}{\left(1+\left( \frac{\left(r-r_i\times f_i\right)}{w_\mathrm{i}}\right)^2\right)}
\end{equation}
Here $A_i$ are the amplitudes of the rings (fit parameters) and N is the number of diffraction rings considered.
For the background function, we use a phenomenological function, which we choose depending on the experimental conditions. Going beyond a phenomenological description of the diffuse background intensity would require knowledge about the phonon properties of the material. For the measurements presented here, we obtained the best agreement to the experimental data using an exponential function plus a linear relationship:
\begin{equation}
    B(r)=a\times\mathrm{exp}\{-r/b\}+c+d\times r
    \label{eq:bg}
\end{equation}
We tested different background functions and found the results for the MSD dynamics to be robust with respect to the choice of background function (less than \unit[8]{\%} deviation in MSD amplitude compared to the results presented here).

Finally, the sum $F+B$ is convolved with a Gaussian to account for the finite coherence of the experiment. The convolution width is a fit parameter. In addition, at this point we add another correction related to the measurement system. Since there are also lens distortions which are not radially symmetric and/or due to spherical aberrations, the outer rings are typically slightly broader in the RP. To account for this, we introduce an additional fit parameter $\delta$ which distorts the radius axis linearly before the convolution:
\begin{equation}
r'=r\times(1-\delta\times\frac{r-r_\mathrm{start}}{r_\mathrm{end}-r_\mathrm{start}})
\end{equation}
$r_\mathrm{start}$ and $r_\mathrm{end}$ are the radii of the beginning and the end of the fit range, respectively. Note that the correction factor $\delta$ depends on the measurement system, in particular on the amount of not radially symmetric distortions and spherical aberration. If these effects are negligible, this correction factor is not necessary and can be set to zero.

\begin{figure}[h]
\begin{center}
\includegraphics[width=0.8\columnwidth]{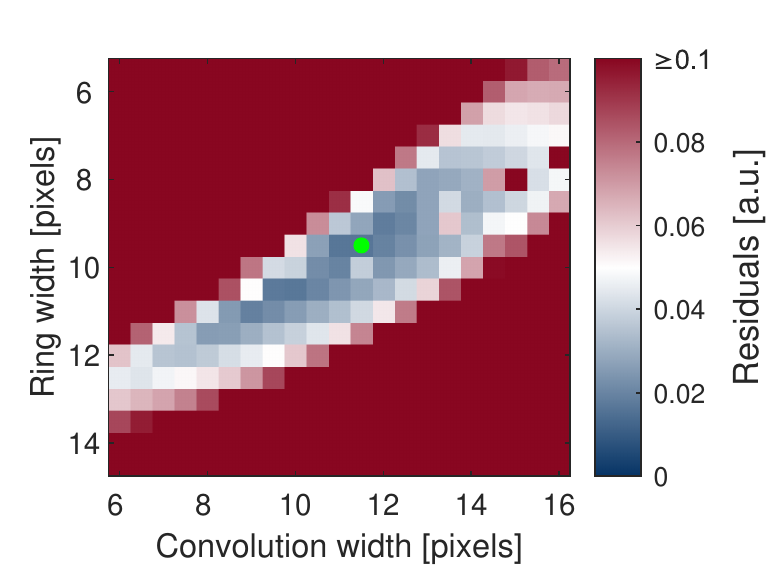}
\caption{Visualization of the width optimization procedure. A map of the fit residuals (sum of squared residuals) as a function of diffraction ring width and convolution width is shown. Both of these parameters were kept fixed in the static fits for width optimization. The combination of widths that yields the lowest residual is marked with a green dot.}
\label{fig:2}
\end{center}
\end{figure}

Finally, it should be noted that the width of the Lorentzians and the Gaussian broadening are strongly correlated. The correlation between them is $\approx -0.48$ when $\delta$ (which influences the average convolution width) is constrained to be zero. Due to this significant correlation, the fit result depends on the starting values. To find the best combination of starting values, the static fit was repeated for all combinations of widths in a reasonable range and with both width parameters fixed. The procedure is similar to a grid search, with the difference being that only the two correlated width parameters form the grid and the other parameters are optimized by the fit. The results are presented in Fig.~\ref{fig:2}.
By this systematic variation of widths, the combination yielding the lowest residual was identified. The static fit was then performed again with this combination of starting parameters. We tested several other starting parameters in the blue region of Fig.~\ref{fig:2} and found the same dynamic fit results within error bars, demonstrating the robustness of our approach.

With our fitting procedure, we obtain excellent agreement to the experimentally measured diffraction pattern, as shown in Figure~\ref{fig:1}(b) and (c). The excellent agreement of the static fit to the experimental data is a very important prerequisite for the second step of the fitting routine, the dynamic fit.

\FloatBarrier
\subsection{Dynamic fit}
In the second step, we analyze changes in the diffraction pattern following laser excitation. In this step, we make use of the results of the static fit, and only allow the following changes to the diffraction pattern compared to the static fit: 
\begin{itemize}
    \item decreases of the diffraction ring intensities corresponding to an MSD increase, according to Debye-Waller theory (see Eq. \ref{eq:Debye-Waller}).
\item changes of the background parameters. Following laser excitation, the background increases due to an increase in diffuse scattering. This effect is also caused by the phonon population increase, see for example Ref.~\cite{2017Waldecker}.
    \item expansion (or contraction) of the sample, leading to a reduction (increase) of the scattering vectors of all rings:
    \begin{equation}
        q_i(t)=q_{i\mathrm{,0}}\times\hspace{2pt}\frac{1}{1+\epsilon(t)}
        \label{eq:expansion}
    \end{equation}
    Here, $q_{i\mathrm{,0}}$ is the scattering vector before laser excitation. Typically, expansion effects are relatively small compared to MSD changes and whether or not expansion is included has no significant effect on the MSD dynamics.

\end{itemize}
Note that this description of the photo-induced changes of the diffraction ring intensities is valid for mono-atomic materials with isotropic MSD, for example mono-atomic face-centered cubic (fcc) or body-centered cubic (bcc) metals. Further information on whether MSDs are isotropic can be found in Ref.~\cite{WillisPryor}. Furthermore, the description can also be applied to many mono-atomic hexagonal close-packed (hcp) metals, depending on their lattice constant ratio~\cite{1971Watanabe}. The description can be extended to more complex materials, based on the crystal structure and the atomic form factors of the constituent ions. 

Especially for low fluences, the noise level of the fit results can be significantly improved by constraining the background parameters, since often the parameters are correlated and local minima are possible. Here we restricted the change in background parameters to no more than 5\% from one delay point to the next. Care was taken to ensure that the fit constraints do not alter the results for the lattice dynamics. An overview of all parameters of the fit function in the static and dynamic fit is presented in Table~\ref{table_parameters}.
\begin{table*}
\vspace{3pt}

\begin{tabular}{L{2cm}|L{5cm}|L{4.7cm}|L{5cm}}
Parameter & Meaning & Static fit & Dynamic fit\\
\hline
\hline
$A_i$     & diffraction ring amplitudes& free parameters & changes only according to Eq.~\ref{eq:Debye-Waller}\\
\hline
$r_1$     & radius of the ring with the lowest scattering vector& free parameter & changes only according to Eq.~\ref{eq:expansion}\\
\hline
$r_i$ (except $i=1$)     & radii of the other rings & determined by their scattering vectors $q_i$, $r_1$, and $\gamma$ & changes only according to Eq.~\ref{eq:expansion}\\
\hline
$f_i$ & radius correction factors & free parameters, but constrained to a few \% around 1 & fixed\\
\hline
$w_\mathrm{reci}$& width of all peaks in reciprocal space (one parameter) & free parameter & fixed \\
\hline
$w_i$& width of the individual peaks in the radial profile & determined by $w_\mathrm{reci}$, $r_i$, and $\gamma$ & fixed \\
\hline
$\gamma$ & correction parameter for radial distortions & free parameter & fixed \\
\hline
convolution width & FWHM of the Gaussian that the calculated pattern is convolved with & free parameter &fixed \\
\hline
$\delta$ & correction parameter for broadening of higher-order peaks  & free parameter & fixed \\
\hline
\hspace{2pt}$\Delta \langle  u^2\rangle $& MSD change & -- & free parameter, determines changes of all $A_i$ simultaneously via Eq.~\ref{eq:Debye-Waller}\\
\hline
$\epsilon$& expansion or contraction & -- & free parameter, determines changes of the lattice constant, i.e. of all $q_i$ simultaneously, via Eq.~\ref{eq:expansion}\\
\hline
background parameters & depends on the background function & free parameters & free parameters (with dynamic constraints). \\
\end{tabular}
\caption{Overview of parameters in the static and dynamic fit.}

\label{table_parameters}
\end{table*}

Figure~\ref{fig:1}(d) presents the differences of the RP's compared to the static RP for several pump-probe delays, as well as the fit results. We obtain good agreement to the experimentally observed changes. The deviations of the ring intensity changes are likely due to multiple scattering effects (within one crystallite of the sample), which lead to deviations from Eq.~\ref{eq:Debye-Waller}. In addition, there are deviations at both ends of the reciprocal space range considered for the analysis, which we attribute to the limitations of the phenomenological background function (see Eq.~\ref{eq:bg}). However, for most of the range considered, both the background and the diffraction rings are well described by the result of the global fit.

\begin{figure*}[bth]
\begin{center}
\includegraphics[width=\textwidth]{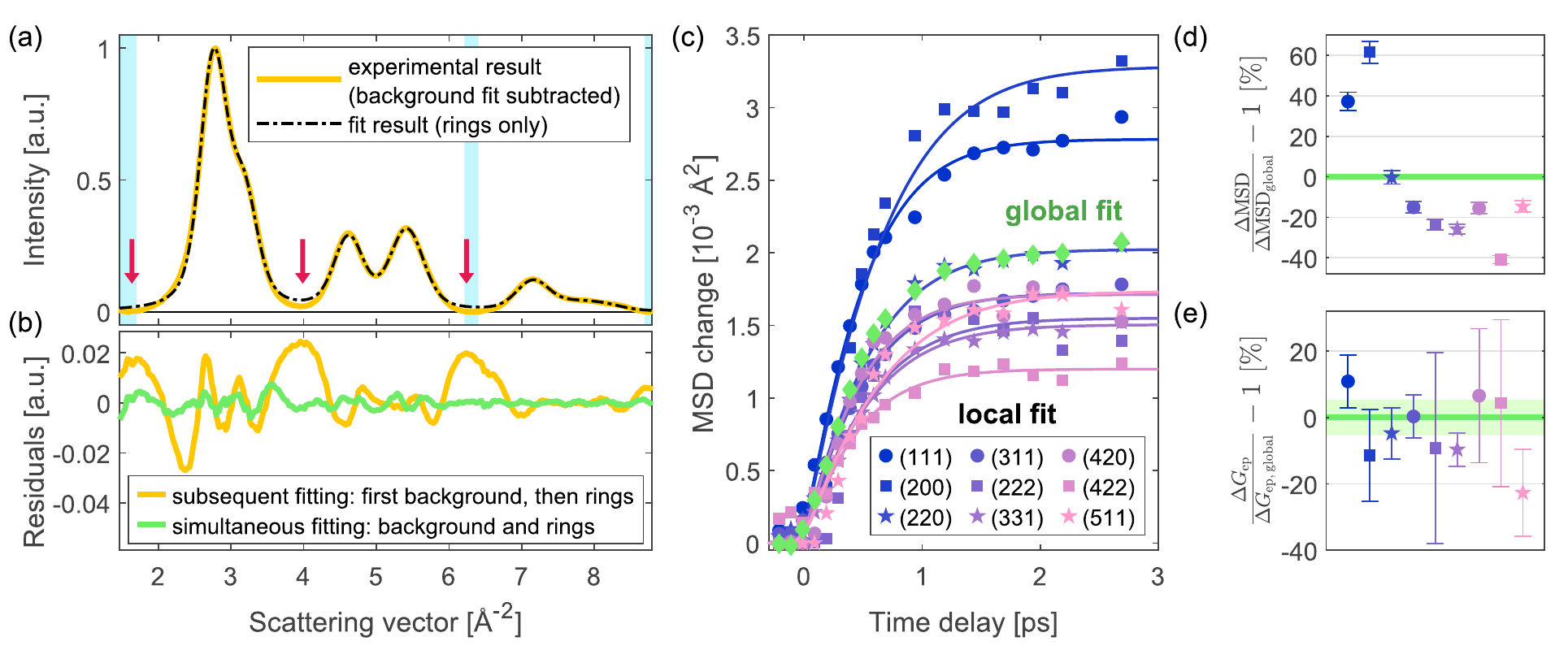}
\caption{Comparison of the global fitting routine with two other analysis methods. (a) Result of a background subtraction before fitting the diffraction rings. The experimental data (yellow) corresponds to the average of all RP's before the arrival of the laser pulse, as used also in the first step of the global fit. The background is subtracted by fitting the background function (see Eq. \ref{eq:bg}) in regions with low intensity from the diffraction rings (shown as light blue areas). This analysis method leads to an overestimation of the background due to residual intensity from the diffraction rings also in the "background" regions. Particularly problematic regions are indicated with red arrows. (b) Comparison of the residuals of this subsequent fitting of background and diffraction rings (yellow curve) to the global fitting procedure, in which both are fitted simultaneously (green curve). The simultaneous fitting of background and rings yields much lower residuals: the root-mean-square deviation (RMSD) of the global fit is only 0.0021, compared to 0.0109 for the subsequent fitting of background and rings. The RMSD units are the same as shown in panel (a), where the first ring is normalized to 1. (c) Result of a local fit of the MSD: The intensity (and thus the MSD change) of each diffraction ring is now a separate fit parameter in the dynamic fit. Results from several diffraction rings are presented, demonstrating that different rings yield different MSD dynamics, particularly in amplitude. The result of the global fit is overplotted. Solid curves are two-temperature model (TTM) fit results.  In all the TTM analyses shown here, the pump laser's arrival time is the same as in the TTM analyses of Section~\ref{sec:4results}. The absorbed energy density of this measurement was \unit[124]{J/cm$^3$}. (d) MSD amplitudes of the individual peaks relative to the global MSD amplitude, calculated from the TTM results at \unit[3]{ps}. The displayed error bars were calculated based on the standard errors from the TTM fits and assuming that the MSD change is proportional to the temperature. The global MSD amplitude ($\unit[0]{\%}$ deviation) is shown as a green line. Its error lies within the line width. (e) Results for the electron-phonon coupling $G_\mathrm{ep}$ from the TTM fits shown in Panel~(c) relative to $G_\mathrm{ep}$ from the global fit. The error bars correspond to the standard errors of the TTM fits. The result of the global analysis is shown as a green line, with the corresponding standard error shown by a shaded area.}
\label{fig:3}
\end{center}
\end{figure*}

In summary, there are two main advantages of the global fitting approach: First, the background and the diffraction rings are fitted together, which allows a reliable background determination. Second, the result for the MSD dynamics is based on the full diffraction pattern instead of individual diffraction rings only. In Fig.~\ref{fig:3}, the global fitting approach is compared to two different analysis methods. Fig.~\ref{fig:3}(a) highlights the first advantage of the global fitting routine by comparing it to the result of a background subtraction and a {\sl subsequent} fit of the diffraction rings. For the latter, in the first step, the background is subtracted by fitting the background function (see Eq.~\ref{eq:bg}) to certain regions in between the diffraction rings, shown as light blue areas. The resulting background-subtracted experimental data is shown as a solid yellow line. In the second step, a fit of the diffraction rings is performed. Here, we use the same fit function as in the global fitting routine, but without any background. The fit result is shown as a dashed black line. The residuals are significantly higher compared to a {\sl simultaneous} fit of background and rings, as shown in Fig.~\ref{fig:3}(b). The finite coherence leads to contributions from the diffraction rings also in the "background" regions, and therefore to an overestimation of the background. An additional disadvantage of this method is that the fit result for the background depends on the choice of "background" regions.

 The second advantage of the global fitting routine is visualized in Fig.~\ref{fig:3}(c)-(e), which show the result if no global MSD change is assumed, but the MSD change is a separate fit parameter for each ring. For this "local fit" of the MSDs we use the same fit function as for the global fit to enable a direct comparison between the two methods. Due to the significant overlap of the individual rings, expansion is not considered here and the MSD of each ring is constrained to be non-negative. The fit was performed analogously to the global fit, hence a fit to the full diffraction pattern was performed (all rings simultaneously, each ring intensity being a separate fit parameter). As shown in Fig.~\ref{fig:3}(c), different diffraction rings yield different results for the MSD change, in particular also significantly different amplitudes of the MSD rise. Regarding the timescale of the MSD rise, no differences are observed within the experimental accuracy. The MSD amplitude differences are illustrated in Fig.~\ref{fig:3}(d), which compares between the result from the global fit and the result from the local fit of individual rings. Often, only a subset of the diffraction rings is considered in analyses of time-resolved diffraction data~\cite{2006Nie,2016Nakamura,Ernstorfer_Au}. Figure~\ref{fig:3}(d) demonstrates that if only one or few rings are considered, the uncertainty in MSD amplitude can potentially be high and the result can depend on which rings are used. In contrast, the global fit uses the full diffraction pattern to extract the MSD dynamics (i.e. {\sl all} rings in the observed range of reciprocal space). Therefore, it removes ambiguities which can arise due to the subjective choice of a subset of diffraction rings for the analysis. Note the systematic dependence on the scattering vector in Figure~\ref{fig:3}(d) suggests that the differences in MSD amplitude could be partially caused by multiple scattering, and we expect smaller discrepancies in the absence thereof. In addition to avoiding a subjective choice of peaks, we expect the minimized number of time-dependent fit parameters and the wide range of reciprocal space considered in the global analysis to be beneficial for the quality of the fit result.

Fig.~\ref{fig:3}(e) presents $G_\mathrm{ep}$ values extracted from the local fit results using a TTM (see Section IV for details on the TTM analysis). The local fit yields different $G_\mathrm{ep}$ values for different diffraction rings and thus an ambiguous result. Here, the deviations from the global fit result are different compared to Fig.~\ref{fig:3}(d) because the result for $G_\mathrm{ep}$ depends on both the amplitude and the timescale of the MSD rise. Nevertheless, also here the comparison between different rings demonstrates that when only a subset of the data is chosen for the analysis, the result can depend on this choice. In contrast, the global fit yields one unambiguous result for the MSD change and thus also one result for $G_\mathrm{ep}$, based on the full diffraction pattern.

\section{Results and discussion}
\label{sec:4results}
\begin{figure}[bth]
\begin{center}
\includegraphics[width=0.9\columnwidth]{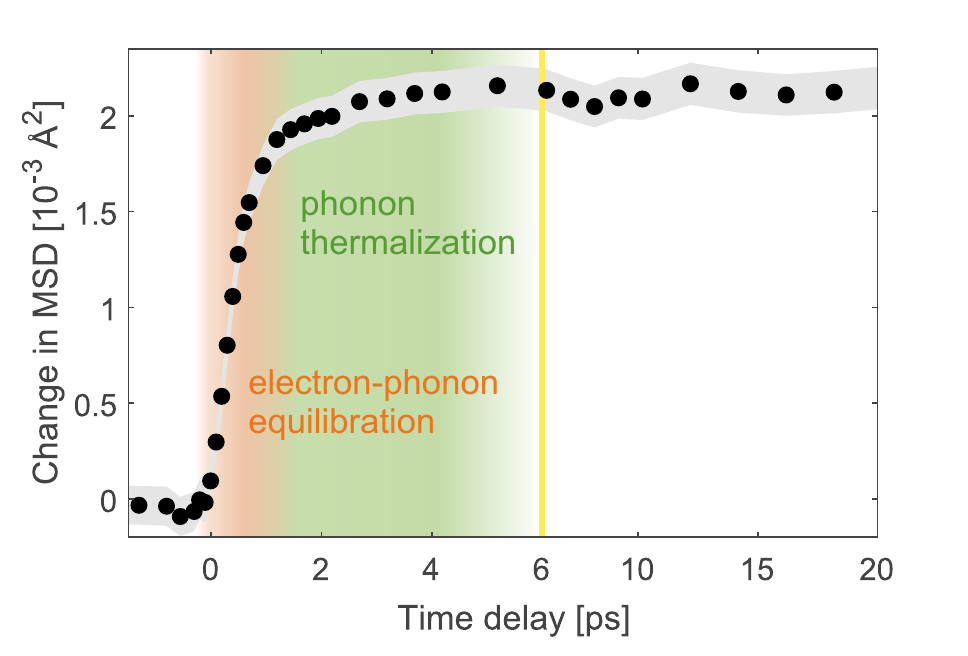}
\caption{Change of atomic mean squared displacement (MSD) as a function of pump-probe delay. The grey shaded area represents the error estimates of the data points, which correspond to the standard deviation obtained from the fitting routine. The yellow line indicates a change of scaling of the time axis. The absorbed energy density of this measurement was \unit[124]{J/cm$^3$}.}
\label{fig:4}
\end{center}
\end{figure}

The MSD evolution of platinum as a function of pump-probe delay extracted with the global fitting routine is presented in Fig.~\ref{fig:4}.
We observe a two-step behavior: A fast component with a time constant of around \unit[600]{fs} and slower, few-picosecond component with a much smaller amplitude. We attribute the fast component to electron-phonon equilibration and the slow component to phonon redistribution processes. Since the second component consists of a further MSD increase, these phonon redistribution processes correspond to energy transfer from higher to lower phonon frequencies, since lower-frequency modes exhibit higher displacements per phonon~\cite{Peng}. In addition, higher-frequency phonons decay into several low-frequency phonons due to their higher energy. Hence, we attribute the slow component of the lattice dynamics to a population increase of low-frequency phonon modes that couple relatively weakly to the electrons and to other phonon modes. Nevertheless, in platinum, the amplitude of the second MSD rise is small compared to the initial MSD rise, which indicates that after the initial rise, most phonon modes have already thermalized with the electrons, except for a small subset of phonons.

\begin{center}
\begin{figure*}[t]

\includegraphics[width=\textwidth]{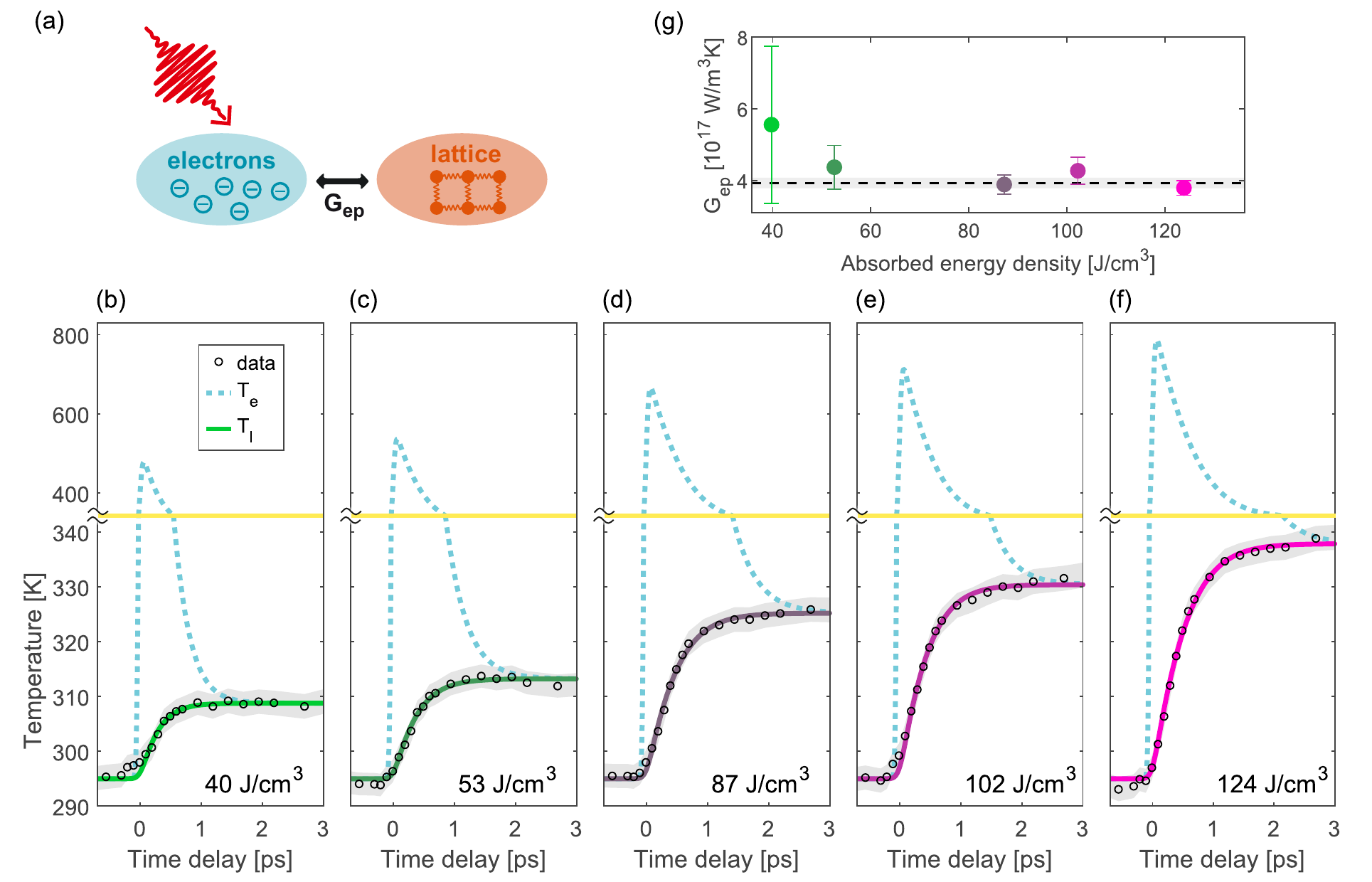}
\caption{Quantitative determination of the electron-phonon coupling constant in platinum using a two-temperature model (TTM). (a) Schematic illustration of the TTM. (b)-(f) Experimental data together with temperature evolution according to the TTM fit results for different absorbed energy densities. The dashed blue curves correspond to the evolution of the electronic temperatures ($T_\mathrm{e}$) and the green to pink curves display the evolution of the lattice temperatures ($T_\mathrm{l}$). The TTM result for the lattice temperature was convolved with a Gaussian (FWHM \unit[150]{fs}) to account for the electron pulse duration (the pump pulse duration is included in the model itself). The yellow line indicates a change of scale in the temperature axis. The experimental results for the lattice temperatures are displayed as black circles and were calculated from the MSD dynamics using the Debye-Waller factor from Ref.~\cite{Peng}. The errors of the experimental data points are displayed as grey shaded areas and correspond to the standard error obtained from the fit of the RP's. (g) TTM fit result for the electron-phonon coupling parameter $G_\mathrm{ep}$ for the absorbed energy densities of Panels~(b-f). The error bars for the $G_\mathrm{ep}$ values correspond to the standard error, which was calculated based on the confidence intervals obtained from the TTM fit. The dashed black line corresponds to the weighted average of the $G_\mathrm{ep}$ values for different absorbed energy densities, and the grey shaded area represents the error of the weighted average.}
\label{fig:5}
\end{figure*}
\end{center}

In the following, in order to study electron-phonon coupling quantitatively, we focus on the initial, fast rise of the MSD, i.e. the time scale from \unit[-1 to 3]{ps}. We convert the MSD rise to lattice temperature using the temperature-dependent Debye-Waller factor provided by Ref.~\cite{Peng}. Based on the results for the lattice temperature, we employ a TTM to model the lattice heating and extract the electron-phonon coupling parameter $G_\mathrm{ep}$. A schematic illustration of the TTM is displayed in Figure~\ref{fig:5}(a).


In the TTM, the material is described as consisting of two heat baths, electrons and phonons, which are always in internal thermal equilibrium. The evolution of the system is described by two coupled differential equations and is governed by the magnitude of $G_\mathrm{ep}$ as well as by the electronic and lattice heat capacities. 

Here, we use the electronic heat capacity calculated by Lin et al.~\cite{2008Lin}. Since platinum is a transition metal, the relationship between electronic heat capacity and temperature is not linear, especially for high electronic temperatures. For electronic temperatures smaller than ca. \unit[700]{K}, the heat capacity calculated by Lin et al. roughly corresponds to $c_e=\gamma \times T $ with \unit[$\gamma\approx 400$]{J/(m$^3$K$^2$)}. This value is in agreement with experimental results for the heat capacity of platinum at room temperature, as discussed in Ref.~\cite{Jang2020}. Here, we don't assume a linear relationship for $c_e$ and directly use the results provided by Lin et al.. For the lattice heat capacity, we use the high-temperature limit derived from equipartition, \unit[24.943]{$\frac{\mathrm{J}}{\mathrm{mol}\hspace{1pt}\mathrm{K}}$} (corresponding to \unit[2.744$\times 10^{6}$]{$\frac{\mathrm{J}}{\mathrm{m}^3\hspace{1pt}\mathrm{K}}$)}\cite{Kittel,Singman}.
This is a valid approximation since the Debye temperature of platinum, \unit[240]{K}\cite{Kittel}, is well below room temperature. 

In the TTM, the laser pulse is assumed to be of Gaussian shape with a FWHM of \unit[80]{fs}. The absorbed energy density and the electron-phonon coupling parameter $G_\mathrm{ep}$ are obtained by finding the best fit of the experimentally measured lattice temperature to the lattice temperature predicted by the TTM. For this fit, the lattice temperature predicted by the TTM is convolved with a Gaussian with a FWHM of \unit[150]{fs} to account for the estimated duration of the electron pulses. Different electron pulse durations (\unit[100]{fs} and \unit[250]{fs}) were also tested and found to have no significant influence on the result for $G_\mathrm{ep}$ (less than \unit[5]{\%} deviation). The arrival time of the laser pulse is the same for all fluences, because all measurements presented here were part of the same data acquisition. To obtain the most accurate value for the arrival time, we first performed TTM fits of the individual measurements with the arrival time as a fit parameter. Then, we calculated the weighted average of the resulting arrival times and repeated the TTM fits with the arrival time fixed at this value.

The evolution of the lattice and electronic temperatures resulting from the TTM fits are displayed in Figure~\ref{fig:5}(b)-(f) for different fluences, together with the experimental results. The TTM fit results yield an excellent description of the experimental data.

The TTM fit result values for $G_\mathrm{ep}$ are shown in Fig.~\ref{fig:5}(g). We don't observe any fluence dependence of $G_\mathrm{ep}$. In principle, based on theoretical calculations, $G_\mathrm{ep}$ is expected to depend on the electronic temperature~\cite{2008Lin}. However, for our range of fluences, a constant $G_\mathrm{ep}$ is a good approximation, since the maximum electronic temperature in our experiments is only around \unit[800]{K} and $G_\mathrm{ep}$ varies only little in that temperature range.

We therefore calculate the weighted average of the $G_\mathrm{ep}$ results for different fluences. The result is $\unit[(3.9\pm0.2)\times10^{17}]{\frac{W}{m^3K}}$. Figure~\ref{fig:5}(g) shows this value as a horizontal dashed line. The error was calculated as the standard error of the weighted mean. Note that it corresponds to the statistical error of $G_\mathrm{ep}$, which was retrieved here in the framework of a TTM and with the DFT calculation results from Ref.~\onlinecite{2008Lin}.

Table~\ref{table1} compares our result for $G_\mathrm{ep}$ to existing literature values obtained from experiments. Our result is within the range of previously measured values. However, note that most literature values for $G_\mathrm{ep}$ were extracted with the electron heat capacity coefficient from low-temperature measurements ($\unit[\gamma\approx750]{\frac{W}{m^3K^2}}$). In contrast, the electron heat capacity we use considers changes of the chemical potential with temperature, which we expect to be more precise at room temperature since the electronic density of states of platinum varies strongly around the Fermi level~\cite{2008Lin}. Since the time evolution of the temperatures in the TTM depends not only on the electron-phonon coupling, but also on the heat capacities, values for $G_\mathrm{ep}$ extracted with different $\gamma$-values are not directly comparable.

The large spread of literature values for $G_\mathrm{ep}$ (in particular also for similar $\gamma$-values) demonstrates that it is non-trivial to extract $G_\mathrm{ep}$ from time-resolved experimental data. Compared to measurements on heterostructures, our experiments have the advantage that the sample is much less complex and no transport effects between different layers occur. In addition, our sample is freestanding, hence there is no carrier and heat transport to a substrate either. Finally, our films are very thin and we probe in transmission, hence transport effects within the platinum layer can also be neglected. Therefore, the lattice response we measure reflects only the intrinsic, microscopic relaxation processes in platinum, which reduces the complexity of extracting  $G_\mathrm{ep}$ from the data.

\begin{table*}
\vspace{3pt}
\begin{tabular}{l|l|l|l}
Authors& $G_\mathrm{ep}$ [$10^{17}$~W/(m$^3$K)]\hspace{3pt}& Method& $\gamma$ [J/(m$^3$K$^2$)]\\
\hline
\hline
Hohlfeld~\cite{2000Hohlfeld} \hspace{3pt}       & 2.5               & TRR                       &740\\
Kimling et al.~\cite{Kimling2017}               & $2.9\pm0.4$       & TRR (het.)                &721\\
{\bf this work}                                 & {\bf 3.9$\pm$ 0.2\hspace{2pt}} & {\bf tr-diffraction}  &$c_e$ from Lin et al.\cite{2008Lin} ($\gamma\approx400$) \\
Choi et al.~\cite{Choi2015}                    & 4.2               & TRR (het.)                &721\\
Jang et al.~\cite{Jang2020}                     & $6\pm1$           & tr-MOKE (het.)            &400\\
Lei et al.~\cite{2002Lei}                      & 6.76              & tr-photoemission         &748\\
Caffrey et al.~\cite{Caffrey2005}               & $10.9\pm0.5$      & TRR                       &750\\
\hline
\end{tabular}
\caption{Comparison of literature values for the electron-phonon coupling parameter $G_\mathrm{ep}$ of platinum obtained from experiments. The third column lists the experimental method that was applied to obtain $G_\mathrm{ep}$. Here, TRR stands for time-resolved reflectivity measurements, "tr" stands for time-resolved, MOKE corresponds to the magneto-optical Kerr effect and "(het.)" indicates that the data was recorded on a heterostructure. The last column lists the value for the electron heat capacity coefficient $\gamma$ that was used to extract $G_\mathrm{ep}$.}
\label{table1}
\end{table*}

The extraction of $G_\mathrm{ep}$ with the TTM is based on the assumption that the two heat baths, electrons and phonons, are always in internal thermal equilibrium. For the electrons, particularly in metals, this is usually a good approximation, since electron-electron scattering is typically more efficient than electron-phonon coupling. For the phonons, a thermal distribution is not always a good approximation on short time scales after excitation~\cite{2017Waldecker,2016Waldecker,SCOPS,2010Trigo,2018Stern,2019Ritzmann,BP1,BP2}. Indeed, also for platinum, we observed signatures of phonon redistribution processes, indicated by the presence of a second, slow MSD rise, as discussed above. However, compared to the initial fast rise of the MSD, the amplitude of the second rise associated with phonon redistribution processes is rather small. For crystals with a trivial basis such as platinum, the MSD caused by a phonon is inversely proportional to its frequency~\cite{Peng}. Due to this strong dependence of the MSD on the phonon frequency, the amount of frequency redistribution corresponding to the second rise is small. Therefore, after the initial electron-phonon equilibration, the phonon frequencies already resemble a Bose-Einstein distribution. In addition, there could however be temperature differences between different phonon modes of the same frequency, which would not necessarily leave signatures in the MSD dynamics. However, if after the initial electron-phonon equilibration there was still a large amount of (weakly coupled) phonon modes with lower temperatures, this would be noticeable as a two-step behavior in the electron dynamics on timescales larger than around \unit[1.5]{ps}. Such a two-step behavior is not observed for thin films of platinum~\cite{Seifert2018,Caffrey2005}. For these reasons, we conclude that for the purpose of describing energy flow from the electrons to the lattice, a TTM is a reasonable approximation. \\
\\

\section{Summary and conclusions}

In this work, we provide a direct measurement of the lattice dynamics of laser-excited platinum using femtosecond electron diffraction. We employ a global fitting routine to extract the changes of atomic mean squared displacement (MSD) reliably from the polycrystalline diffraction patterns, which we describe in detail. The approach can be applied to all mono-atomic materials with isotropic MSD, and could also be further extended to describe more complex materials or heterostructures. We extract the MSD evolution of platinum following laser excitation, which exhibits two time scales: a sub-picosecond MSD rise, which we attribute to electron-phonon equilibration, and a further, much smaller MSD rise on a few-picosecond time scale, which we attribute to phonon-phonon redistribution processes. 
Based on the dominant, fast MSD rise and using a two-temperature model (TTM), we extract a value of $\unit[(3.9\pm0.2)\times10^{17}]{\frac{W}{m^3K}}$ for the electron-phonon coupling parameter $G_\mathrm{ep}$. Within the range of fluences applied in our experiment, we don't observe any fluence dependence of $G_\mathrm{ep}$. Compared to previous reports of $G_\mathrm{ep}$, our approach has the advantage that our sample is a freestanding thin film, hence transport effects don't play a role in the dynamics. Furthermore, in contrast to optical spectroscopy, our technique is sensitive only to one subsystem, the lattice. We expect that precise knowledge of electron-phonon coupling in platinum will benefit the modeling and understanding of heterostructures containing this material, for example spintronic devices and photocatalytic structures.

\section*{Data availability}
The data that support the findings of this study are available on a data repository~\cite{Pt_data}. The code of the global fitting routine is also available~\cite{Pt_code}.
\\

\section*{Acknowledgement}
We thank Thomas Vasileiadis and Reza Rouzegar for helpful discussions and Sven Kubala for sample growth. This work was funded by the Deutsche Forschungsgemeinschaft (DFG) through SFB/TRR 227  "Ultrafast Spin Dynamics" (Project B07) and through the Emmy Noether program under Grant No. RE 3977/1, by the European Research Council (ERC) under the European Union’s Horizon 2020 research and innovation program (Grant Agreement Number ERC-2015-CoG-682843), and by the Max Planck Society. H.S.~acknowledges support by the Swiss National Science Foundation under Grant No.~P2SKP2\textunderscore184100.

\bibliography{main}

\end{document}